\documentclass[aps,prl,twocolumn,groupedaddress,showpacs]{revtex4}
\usepackage{graphics}

\usepackage[]{graphicx}
\DeclareGraphicsExtensions{.pdf}

\newcommand {\e}[1]{\mathrm{~#1}}       % za pisanje enot
\newcommand {\E}[1]{\cdot 10^{#1}}		% za pisanje desetiskih eksponentov								
\newcommand {\vek}[1]{\mathbf{#1}}

\begin{document}

\title{Ising models for networks of real neurons}

\author{Ga\v{s}per Tka\v{c}ik,$^{a,c}$ Elad Schneidman,$^{a-c}$ Michael J. Berry II,$^{b}$ and William Bialek$^{a,c}$}
\affiliation{$^a$Joseph Henry Laboratories of Physics, $^b$Department of Molecular Biology, and
$^c$Lewis--Sigler Institute for Integrative Genomics\\
Princeton University, Princeton, New Jersey 08544 USA}

\date{\today}

\begin{abstract}
Ising models with pairwise interactions are the least structured, or maximum--entropy, probability distributions that exactly reproduce measured pairwise correlations between spins. Here we use this equivalence to construct Ising models that describe the correlated spiking activity of populations of 40 neurons in the retina, and show that pairwise interactions account for observed higher--order correlations. By first finding a representative ensemble for observed networks we can create synthetic networks of 120 neurons, and find that with increasing size the networks operate closer to a critical point and start exhibiting collective behaviors reminiscent of spin glasses.
\end{abstract}

% neural networks, info processing in vision & hearing, info theory, 05.20.-y classical statmech
%\keywords{entropy, information, multi--information, neural networks, Monte Carlo, correlation}
\maketitle

Physicists have long explored analogies between the statistical mechanics of Ising models and the functional dynamics of neural networks \cite{hopfield_82,amit_89}.
Recently it has been suggested that this analogy can be turned into a  precise mapping \cite{schneidman+al_06}:  In small windows of time,  a single neuron ${\rm i}$ either does ($\sigma_{\rm i}=+1$) or does not ($\sigma_{\rm i} = -1$) generate an action potential or ``spike'' \cite{spikes}; if we  measure the mean probability of  spiking for each cell ($\langle \sigma_{\rm i}\rangle$) and the correlations between pairs of cells ($C_{\rm ij} = \langle \sigma_{\rm i}\sigma_{\rm j} \rangle  -  \langle \sigma_{\rm i}\rangle\langle\sigma_{\rm j} \rangle$), then the maximum entropy model consistent with these data is {\sl exactly} the Ising model
\begin{equation}
P(\{\sigma_{\rm i}\}) = {1\over{Z}}\exp\left[
\sum_{{\rm i}=1}^N h_{\rm i}\sigma_{\rm i} + {1\over 2}\sum_{{\rm i}\neq {\rm j}}^N J_{\rm ij} \sigma_{\rm i}\sigma_{\rm j} \right] ,
\end{equation}
where the magnetic fields $\{h_{\rm i}\}$ and the exchange couplings $\{J_{\rm ij}\}$ have to be set to reproduce the measured values of $\{\langle \sigma_{\rm i}\rangle\}$ and $\{C_{\rm ij}\}$.
We recall that maximum entropy models are the least structured models consistent with known expectation values \cite{jaynes_57,schneidman+al_03}; thus the Ising model
  is the minimal model forced upon us by measurements of mean spike probabilities and pairwise correlations.  

The surprising result of Ref \cite{schneidman+al_06} is that the Ising model provides a very accurate description of the combinatorial patterns of spiking and silence in retinal ganglion cells as they respond to natural movies, despite the fact that the model explicitly discards all higher order interactions among multiple cells.  This detailed comparison of theory and experiment was done for groups of $N\sim 10$ neurons, which are 
small enough that the full distribution $P(\{\sigma_{\rm i}\})$ can be sampled experimentally.
Here we extend these results to $N=40$, and then argue that the observed network is typical of an ensemble out of which we can construct larger networks.
 Remarkably, these larger networks seem to be  poised very close to a critical point, and exhibit other collective behaviors which should become visible in the next generation of experiments.

To be concrete, we consider the salamander retina responding to naturalistic movie clips, as in the experiments of Refs \cite{schneidman+al_06,puchalla+al_05}.  Under these conditions, pairs of cells within  $\sim 200\,\mu{\rm m}$ of each other have correlations drawn from a homogeneous distribution; the correlations decline at larger distance
\cite{approx}.   This correlated patch contains $N\sim 200$ neurons, of which we record from $N=40$ \cite{expts}; experiments typically run for $\sim 1\,{\rm hr}$ \cite{data}.

The central problem is to find the magnetic fields and exchange interactions that reproduce the observed pairwise correlations.  It is convenient to think of this problem more generally:  We have a set of operators $\hat O_\mu (\{\sigma_{\rm i}\})$ on the state of the system, and we consider a class of models 
\begin{equation}
P(\{\sigma_{\rm i}\}| {\vek g}) = {1\over{Z({\vek g})}}
\exp\left[ \sum_{\mu =1}^K g_\mu \hat  O_\mu (\{\sigma_{\rm i}\}) \right] ;
\end{equation}
our problem is to find the coupling constants $\vek g$ that generate the correct expectation values, which is equivalent to solving the equations
$
{{\partial \ln Z({\vek g})} /{\partial g_\mu}} = \langle \hat O_\mu (\{\sigma_{\rm i}\}) \rangle_{\rm expt} .
$
Up to $N\sim20$ cells we can solve  exactly, but
this approach does not scale to $N=40$ and beyond.  For larger systems, this ``inverse Ising problem'' or Boltzmann machine learning, as it is known in computer science \cite{Hinton}, is a hard computational problem rarely encountered in physics, where we usually compute properties of the system given  a known  model of the interactions. 

Given a set of coupling constants $\vek g$, we can estimate the expectation values 
$\langle\hat O_\mu \rangle_{\vek g}$ by Monte Carlo simulation.  Increasing the  coupling $g_\mu$ will increase the expectation value $\langle\hat O_\mu \rangle$, so a plausible algorithm for learning   $\vek g$ is to increase each $g_\mu$ in proportion to the deviation of $\langle\hat O_\mu \rangle$ (as estimated by Monte Carlo) from its target value (as estimated from experiment).  This is not a true gradient ascent, since changing $g_\mu$ has an impact on operators $\langle\hat O_{\nu\neq\mu} \rangle$, but such an iteration scheme does have the correct fixed points; heuristic improvements 
include a slowing of the learning rate with time and the addition of some `inertia', so that we update $g_\mu$ according to 
\begin{equation}
\Delta g_{\mu}(t+1) = - \eta(t) \left[
\langle\hat{O}_\mu\rangle_{{\vek g}(t)} 
-
\langle\hat{O}_\mu  \rangle_{\rm expt} 
\right]
+\alpha \Delta g_{\mu}(t),
\label{eq_ising_learn}
\end{equation}
where $\eta(t)$ is the time--dependent learning rate and $\alpha$ measures the strength of the inertial term \cite{mcmc}.

\begin{figure} %  figure placement: here, top, bottom, or page
   \centering
   \includegraphics[width=3.5in]{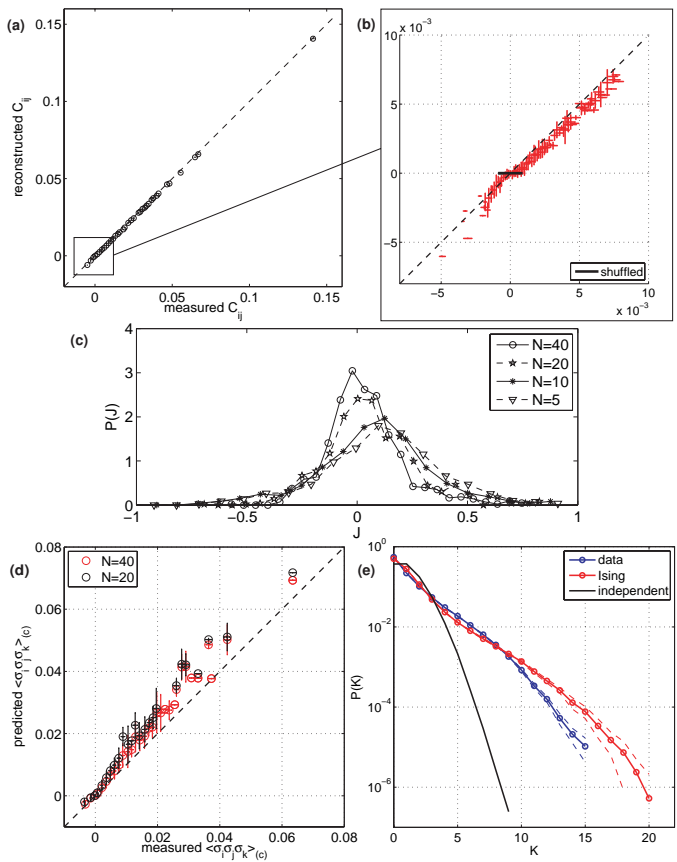}
   \caption{(a) Precision of the Ising model learned via  Eq (\ref{eq_ising_learn}): measured covariance elements are binned on the x-axis and plotted against the corresponding reconstructed covariances on y-axis; vertical error bars denote the deviation within the bin and horizontal error bars denote the bootstrap errors on covariance estimates.  (b) Zoom--in for small $C_{\rm ij}$,  with scale bar representing the distribution of covariances from shuffled data. Not shown are the reconstructions of the means, $\langle\sigma_{\rm i}\rangle$, which are accurate to better than $1\%$. (c) Distribution of coupling constants $J_\mathrm{ij}$. (d) Measured vs predicted connected three--point correlations for 40 neurons (red) and exact solution for a 20 neuron subset (black). (e) Probability of observing $K$ simultaneous spikes, compared to the failure of the independent model (black line). Dashed lines show error estimates.} 
   \label{f1}
\end{figure}

Figure \ref{f1} shows the success of the learning algorithm by comparing the measured pairwise correlations to those computed from the inferred Ising model for 40 neurons.
To verify that the pairwise Hamiltonian captures essential features of the data, we predict and then check statistics that are sensitive to higher order structure: the probability $P(K)$ of patterns with $K$ simultaneous spikes, connected triplet correlations and the distribution of energies 
(latter not shown). The model overestimates the significant 3--point correlations by about $7\%$ and generates small deviations in $P(K)$;  most notably it underestimates the no--spike pattern, $P_{\rm expt}(K=0) =0.550$ vs. $P_{\rm Ising} (K=0) =0.502$. These deviations are small, however, and it seems fair to conclude that the pairwise Ising model captures the structure of the $N=40$ neuron system very well.  Smaller groups of neurons for which exact pairwise models are computable also show excellent agreement with the data \cite{schneidman+al_06, deviations}.

It is  surprising that pairwise models work well both on $N=40$ neurons and on smaller subsets of these: not observing $\sigma_\chi$ will induce a triplet interaction among neurons $\left\{\sigma_\alpha,\sigma_\beta,\sigma_\gamma\right\}$ for any triplet in which there were pairwise couplings between $\sigma_\chi$ and all triplet members. 
Moreover, comparison of the parameters in $\vek{g}^{(40)}$ with their corresponding averages from different subnets $\vek{g}^{(20)}$ leaves the exchange interactions almost unchanged, while magnetic fields change substantially.  To explain both phenomena, we 
examine the flow of the couplings under decimation.  Specifically, we include three--body interactions, isolate terms related to spin $\sigma_{\rm n}$, sum over $\sigma_{\rm n}$, expand in $J_{\rm in},J_{\rm ijn}$ up to  $O(\sigma^4)$, and then identify renormalized couplings:
\begin{eqnarray}
h_{\rm i}&\rightarrow & h_{\rm i} + \omega \tilde{J}_{\rm in} +\textstyle \sum_{\rm j} \beta_{\rm ij}+{\cal O}(\gamma,\delta),	\label{eq_dec1}\\
J_{\rm ij}&\rightarrow & J_{\rm ij}+\beta_{\rm ij}+{\cal O}(\gamma,\delta),\label{eq_dec2} \\
J_{\rm ijk} &\rightarrow& J_{\rm ijk} + {\cal O}(\gamma,\delta) \label{eq_dec3}
\end{eqnarray}
where $\tilde{J}_{\rm in}=J_{\rm in}-\sum_{\rm j} J_{\rm ijn}$, $\beta_{\rm ij} = \tilde{J}_{\rm in}\tilde{J}_{\rm jn}(1-\omega^2)+\omega J_{\rm ijn}$ and  $\omega=\tanh(h_{\rm n}-\sum_{\rm i} J_{\rm in}+\frac{1}{2}\sum_{\rm ij} J_{\rm ijn})$.  The terms  $\gamma,\delta \propto (1-\omega^2)$ originate from terms with 3 and 4 factors of $\sigma$, respectively.
The key point is that neurons spike very infrequently (on average in $\sim 2.4\%$ of the bins) and so $\langle \sigma_{\rm i}\rangle\approx-1$, in which case $\omega$ is approximately the hyperbolic tangent of the mean field at site $\rm n$ and is close to $-1$. If pairwise Ising  is a good model at size $N$, and couplings are small enough to permit expansion, then at size $(N-1)$ the corrections to pairwise terms, as well as $J_{\rm ijk}$, are suppressed by $1-\omega^2$.  This  could explain the dominance of pairwise interactions:  it is not that higher order terms are intrinsically small, but the fact that spiking is rare means that they do not have much chance to contribute.  Thus, the pairwise approximation is more like a Mayer cluster or virial expansion than like simple perturbation theory.

We test these ideas by selecting 100 random subgroups of 10 neurons out of  20; for each, we compute the exact Ising model from the data, as well as 
applying 
Eqs (\ref{eq_dec1}--\ref{eq_dec3})
10 times in succession to decimate the network from 20 cells down to the chosen 10. The resulting three--body interactions $J_{\rm ijk}$ have a mean and standard deviation ten times smaller than the pairwise $J_{\rm ij}$.  If we ignore these terms,  the average Jensen--Shannon divergence \cite{lin_91} between this probability distribution and the best pairwise model for the $N=10$  subgroups is $\overline{D}_{JS}= 9.3\pm 5.4 \times 10^{-4}\,{\rm bits}$, which is smaller than the average divergence between either model and the experimental data and means that $\gg10^3$  samples would be required to distinguish reliably between the two models. Thus, sparsity of spikes keeps the complexity in check.

Given a model with couplings $\vek g$, we can explore the statistical mechanics of models with ${\vek g} \rightarrow {\vek g}/T$.  In particular, this exercise might reveal if the actual operating point ($T=1$) is in any way privileged.
Tracking the specific heat vs $T$ also gives us a way of estimating the entropy at $T=1$, which measures the capacity of the neurons to convey information about the visual world; we recall that   
$S(T=1) =\int_0^1 C(T)/T\,dT$, and the heat capacity  can be estimated by Monte Carlo from the variance of the energy, $C(T)=\langle (\delta E)^2\rangle/T^2$. 

 \begin{figure} %  figure placement: here, top, bottom, or page
   \centering
   \includegraphics[width=3.5in]{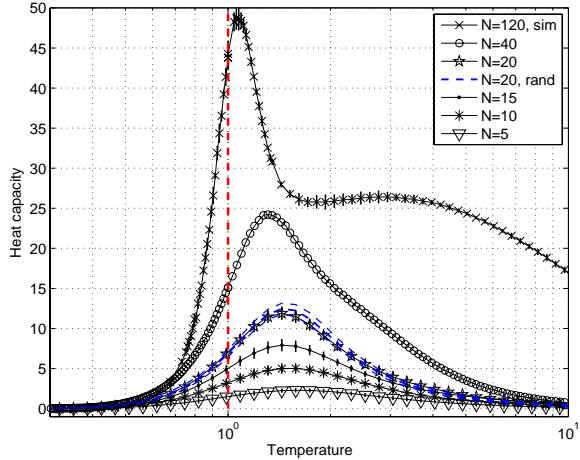}
   \caption{$C(T)$ for systems of different sizes. Ising models were constructed for 400 subnetworks of size 5, 180 of size 10, 90 of size 15 and 20, 1 full network of size 40 (all from data), and 3 synthetic networks of size 120; vertical error bars are standard deviations across these examples. The mean of the heat capacity curve and the 1 sigma envelope for Ising models of randomized networks are shown in blue dashed lines. 
   } 
   \label{f2}
\end{figure}

Figure \ref{f2} shows the dependence of heat capacity on temperature at various system sizes.  
We note that the peak of the heat capacity moves towards the operating point with increasing size.
The behavior of the heat capacity $C(T)$ is diagnostic for the underlying density of states, and offers us the chance to ask if the networks we observe in the retina are typical of some statistical ensemble.  One could generate such an ensemble by randomly choosing the matrix elements $J_{\rm ij}$ from the distribution that characterizes the real system, but models generated in this way have wildly different values of $\langle \sigma_{\rm i}\rangle$.  An alternative is to consider that these expectation values, as well as the pairwise correlations $C_{\rm ij}$, are drawn independently out of a distribution, and then we construct Ising model consistent with these randomly assigned expectation values.  Figure \ref{f2} shows $C(T)$ for networks of 20 neurons constructed in this way \cite{rshuffle}, and we see that, within error bars, the behavior of these randomly chosen systems resembles that of real 20 neuron groups in the retina.

Armed with the results at $N=20$, we generated several
synthetic networks of 120 neurons by randomly choosing once more out of the distribution of $\langle\sigma_{\rm i}\rangle$ and $C_{\rm ij}$ observed experimentally \cite{bignet}.  The heat capacity $C_{120}(T)$ now has a dramatic peak at
$T^*=1.07\pm0.02$, very close to the operating point at $T=1$.  If we integrate to find the entropy, we find that   the independent entropy of the individual spins, $S_0 (120) = 17.8\pm 0.2\,{\rm bits}$,  has been reduced to $S(120) = 10.7\pm 0.2\,{\rm bits}$.  Even at $N=120$ the entropy deficit or multi--information $I(N) = S_0(N) - S(N)$ continues to grow in proportion to the number of pairs ($\sim N^2$), continuing the pattern found in smaller networks \cite{schneidman+al_06}. Looking in detail at the model,   the distribution of   $J_{\rm ij}$ is approximately Gaussian  $\overline{J}=-0.016\pm 0.004$ and $\sigma_J=0.61\pm 0.04$;
$53\%$ of triangles are frustrated ($46\%$ at $N=40$), indicating the possibility of many  stable states, as in spin glasses \cite{mezard+al_87}.  We examine these next.

\begin{figure}[b] %  figure placement: here, top, bottom, or page
   \centering
   \includegraphics[width=3in]{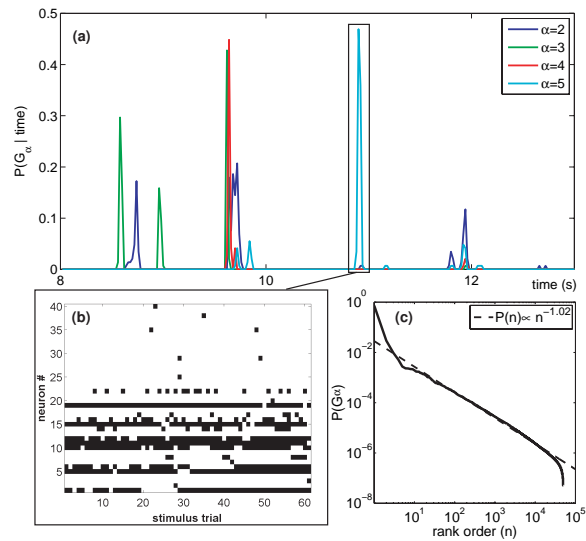}
   \caption{(a) 
   Probability that the 40 neuron system is found in a configuration within the basin of each nontrivial ground state ${\cal G}_{\alpha}$, as a function of time during the stimulus movie; $P({\cal G}_{\rm \alpha} | t) = 0.4$ means that the retina is in that basin on $40\%$ of the 145 repetitions of the movie.
   (b)   All unique patterns  assigned to $\mathcal{G}_5$ at $t=10.88-10.92\e{s}$. (c) Zipf plot of the rank ordered frequencies with which the lowest lying $5\E{4}$ stable states are found in the simulated 120 neuron system.
   }
   \label{f4}
\end{figure}

At $N=40$ we find 4  
local energy minima ($\mathcal{G}_2,\cdots ,\mathcal{G}_5$) in the observed sample that are stable against single spin flips, in addition to the silent state $\mathcal{G}_1$ ($\sigma_{\rm i}=-1$ for all $\rm i$). Using zero--temperature Monte Carlo, each configuration observed in the experimental data is assigned to 
its corresponding stable state.   Although this assignment makes no reference to the visual stimulus,  the  collective states ${\cal G}_\alpha$ are reproducible across multiple presentations of the same movie (Fig \ref{f4}a), even when the microscopic state $\{\sigma_{\rm i}\}$ varies substantially (Fig \ref{f4}b).

At $N=120$,  we find a much richer structure \cite{longmmc}: the Gibbs state now is a superposition of  thousands of $\mathcal{G}_\alpha$, with a nearly Zipf--like distribution (Fig \ref{f4}c).  The entropy of this distribution is $3.4\pm 0.3 \,{\rm bits}$, about a third of the total entropy.  Thus, a substantial fraction of the network's capacity to convey visual information would be carried by the collective state, that is by the identity of the basin of attraction, rather than by the detailed microscopic states.

To summarize,  the Ising model with pairwise interactions continues to provide an accurate description of neural activity in the retina up to $N=40$.  Although correlations among pairs of cells are weak, the behavior of these large groups of cells is strongly collective, and this is even clearer in larger networks that were constructed to be typical of the ensemble out of which the observed network has been drawn.  In particular, these networks seem to be operating close to a critical point.  Such tuning might serve to maximize the system's susceptibility to sensory inputs, as  suggested in other systems \cite{critical}; by definition operating at a peak of the specific heat maximizes the dynamic range of log probabilities for the different microscopic states,  allowing the system to represent sensory events that occur with a wide range of likelihoods \cite{contrast}.  The observed correlations are not fixed by the anatomy of the retina or by the visual input alone, but reflect adaptation   to the statistics of these inputs \cite{adapt}; it should be possible to test experimentally whether these adaptation processes preserve the tuning to a critical point as the input statistics are changed.  Finally, the transition from $N=40$ to $N=120$ opens up a much richer structure to the configuration space, suggesting that the representation of the visual world by the relevant groups of $N\sim 200$ cells may be completely dominated by collective states that are invisible to experiments on smaller systems.

\acknowledgments {This work was supported in part by NIH  Grants R01 EY14196 and P50 GM071508, by the E. Matilda Zeigler Foundation, by NSF Grant IIS--0613435 
and by the Burroughs Wellcome Fund Program in Biological Dynamics.}

%
%
%
%
%
%
%
%
%
%
%%%%%%%%%%%%%%%%%%%%%%%%%

%

\begin{thebibliography}{99}
\bibitem{approx}
These are approximate statements; for details see \cite{puchalla+al_05}.
\bibitem{expts}
Alternative recording methods can capture more cells at low density \cite{litke+al_04}, or fewer cells at higher density \cite{segev+al_04}.
\bibitem{data}
Some experimental details \cite{schneidman+al_06}:
The visual stimulus consists of a $26.2\e{s}$ movie that was projected onto the retina 145 times in succession; using $\Delta \tau=20\e{ms}$ quantization this yields 1310 samples per movie repeat, for a total of 189950 samples. The effective number of independent samples is smaller because of correlations across time; using bootstrap error analysis we estimate $N_{\rm samp} \sim 7\E{4}$.
\bibitem{mcmc}
The learning rate $\eta(t)$ was decreased as $O(1/t)$ or slower according to a custom schedule; $\alpha=0.9$ for a network of $N=120$ neurons and 0 otherwise. An  initial approximate solution for $\vek{g}$ was obtained by contrastive divergence (CD) Monte Carlo \cite{cd} for 40 neurons for which we have the complete set of patterns needed by CD.  The Hamiltonian was rewritten such that $J_{\rm ij}$ was constraining $(\sigma_{\rm i}-\langle\sigma_{\rm i}\rangle_{\rm expt})(\sigma_{\rm j}-\langle\sigma_{\rm j}\rangle_{\rm expt})$, and we found that this removed   biases in the reconstructed covariances.
\bibitem{deviations}
Exact pairwise models at $N=20$ exhibit similar but smaller systematic deviations from the data, suggesting that the deviations are not due to convergence problems.
\bibitem{rshuffle}
Not all combinations of means and correlations are possible for Ising variables.
After each draw from the distribution of $\langle \sigma_{\rm i}\rangle$ and $C_{\rm ij}$, we check that all 
$2\times 2$ marginal distributions are in $\left[0,1\right]$, and repeat if needed. Once the whole synthetic covariance matrix is generated, we check (e.g. using Kolmogorov--Smirnov) that the distribution of its elements is consistent with the measured distribution.
\bibitem{bignet}
Learning the couplings $\vek g$ was slow, but eventually converged: $C_{\rm ij}$ converged to within $10\%$ for the largest quartile of elements by absolute value,
and within $15\%$ for the largest half,
without obvious systematic biases.
\bibitem{longmmc}
One run of $2\E{7}$ independent samples ($\tau_\mathcal{G}^{-1}\approx 5\E{2}\mathrm{flips}$) is collected; for each sample ZTMC is used to determine the appropriate basin; we track $5\E{4}$ lowest energy stable states and keep detailed statistics for $10^3$ lowest.
\bibitem{contrast}
In contrast,  simple notions of efficient coding require all symbols to be used with equal probability.  But since states of the visual world occur with wildly varying probabilities, this (and related notions of efficiency) require codes that are extended over time.  If  the brain is interested directly in how surprised it should be by the current state of its inputs, then it might be more important to maximize the dynamic range for instantaneously representing this (negative log) likelihood.
\end{thebibliography}
 \end{document}